\documentclass[11pt]{article}
\usepackage{float}
\usepackage{times}
\usepackage{amsmath}
\usepackage{graphicx,graphics}
\usepackage{bm,bbm }
\usepackage{authblk}
\usepackage{epstopdf,epsfig,subfigure}
\usepackage{amsmath,amssymb,amsfonts}
\def\UnderWiggleTemp{\the\catcode`\@}
\catcode`\@=11
\ifx\UnderWiggle@Loaded\relax
  \catcode`\@=\UnderWiggleTemp
    \else \let\UnderWiggle@Loaded=\relax \fi
\newbox\U@BoxA
\newbox\U@BoxB
\newdimen\U@DimenA
\def\U@DoUnderWiggle{
  \offinterlineskip
  \vtop{
    \hbox{\vbox{\copy0}}
    \vskip 1.2pt  
    \vbox to 0.4pt{
      \hbox to\wd0{\hss\char'176\hss}
      \vskip0pt minus 1fil
    }
    \vskip 0.4pt  
  }
}
\def\UnderWiggle#1{{%
  \ifmmode
    \mathchoice
      {\setbox0=\hbox{$\displaystyle #1$}\U@DoUnderWiggle}
      {\setbox0=\hbox{$\textstyle #1$}\U@DoUnderWiggle}
      {\setbox0=\hbox{$\scriptstyle #1$}\U@DoUnderWiggle}
      {\setbox0=\hbox{$\scriptscriptstyle #1$}\U@DoUnderWiggle}
  \else
    \setbox0=\hbox{#1}\U@DoUnderWiggle
  \fi
}}
\catcode`\@=\UnderWiggleTemp

\newcommand{\be}{\begin{eqnarray}}         
\newcommand{\ee}{\end{eqnarray}}           \newcommand{\ba}{\begin{eqnarray*}}
\newcommand{\ea}{\end{eqnarray*}}   \newcommand{\eaa}{\end{align}}                 
\newcommand{\baa}{\begin{align}}

\newcommand{\bl}{\begin{lemma}}
\newcommand{\el}{\end{lemma}}              \newcommand{\bd}{\begin{definition}}
                 \newcommand{\ed}{\end{definition}}
               
 \newcommand{\bc}{\begin{corollary}}
\newcommand{\ec}{\end{corollary}}          

     \newcommand{\var}{\mbox{Var}}

\newtheorem{theorem}{Theorem}[section]     \newtheorem{corollary}{Corollary}[section]
\newtheorem{lemma}{Lemma}[section]         \newtheorem{remark}{Remark}[section]
\newtheorem{example}{Example}[section]     \newtheorem{definition}{Definition}[section]

\def\boldfacefake #1{%
    \hbox{%
        \mathsurround=0pt
        \hbox to 0.4pt{$#1$\hss}%
        \hbox to 0.4pt{$#1$\hss}%
        \hbox {$#1$}%
    }%
}

\newcommand{\twof}[4]
{ \hbox to\hsize{\hss
    \vbox{\psfig{figure=\mydir/#1,width=#3,height=#4}}\qquad
    \vbox{\psfig{figure=\mydir/#2,width=#3,height=#4}}
    \hss}
\vskip -0.0truein \hbox to\hsize{\hss
    \vbox{ \begin{center}\mbox{\footnotesize \hspace{0.0in} {(a)}
                     \hspace{#3} {(b)}  }  \end{center} }
    \hss}
\vskip 0.0truein }
\newcommand{\threefig}[5]
{ \hbox to\hsize{\hss
  \vbox{\psfig{figure=\mydir/#1,width=#4,height=#5}}%
  \hss}
\hbox to\hsize{\hss
  \vbox{\psfig{figure=\mydir/#2,width=#4,height=#5}}%
  \hss}
\hbox to\hsize{\hss
  \vbox{\psfig{figure=\mydir/#3,width=#4,height=#5}}%
  \hss}
}
\newcommand{\threefigv}[5]
{ \hbox to\hsize{\hss
  \vbox{\psfig{figure=\mydir/#1,width=#4,height=#5}}%
  \hss}
  \vskip -0.1truein
\hbox to\hsize{\hss
    \vbox{\begin{center} \footnotesize{(a)} \end{center}}
    \hss}
\vskip -0.1truein \hbox to\hsize{\hss
  \vbox{\psfig{figure=\mydir/#2,width=#4,height=#5}}%
  \hss}
  \vskip -0.1truein
\hbox to\hsize{\hss
    \vbox{\begin{center} \footnotesize{(b)} \end{center}}
    \hss}
\vskip -0.1truein \hbox to\hsize{\hss
  \vbox{\psfig{figure=\mydir/#3,width=#4,height=#5}}%
  \hss}
  \vskip -0.1truein
\hbox to\hsize{\hss
    \vbox{\begin{center} \footnotesize{(c)} \end{center}}
    \hss}
}
\newcommand{\threesmall}[5]{
\hbox to\hsize{\hss
    \vbox{\psfig{figure=\mydir/#1,width=#4,height=#5}} \hspace{0.2in}
    \vbox{\psfig{figure=\mydir/#2,width=#4,height=#5}} \hspace{0.2in}
    \vbox{\psfig{figure=\mydir/#3,width=#4,height=#5}}
    \hss}
\vskip -0.1truein \hbox to\hsize{\hss
    \vbox{ \begin{center}\mbox{\footnotesize \hspace{0.0in} {(a)}
                     \hspace{#4} {(b)}   \hspace{#4} {(c)}}  \end{center} }
    \hss}
}
\newcommand{\fourf}[6]
{ \hbox to\hsize{\hss
    \vbox{\psfig{figure=\mydir/#1,width=#5,height=#6}}\qquad
    \vbox{\psfig{figure=\mydir/#2,width=#5,height=#6}}
    \hss}
\vskip 0.1truein \hbox to\hsize{\hss
    \vbox{ \begin{center}\mbox{\footnotesize \hspace{0.1in} {(a)}
                     \hspace{#5} {(b)}  }  \end{center} }
    \hss}
\vskip 0.1truein \hbox to\hsize{\hss
    \vbox{\psfig{figure=\mydir/#3,width=#5,height=#6}}\qquad
    \vbox{\psfig{figure=\mydir/#4,width=#5,height=#6}}
    \hss}
\vskip -0.1truein
    \vbox{ \begin{center}\mbox{\footnotesize \hspace{0.1in} {(c)}
                     \hspace{#5} {(d)}  }  \end{center} }
\hbox to\hsize{\hss
    \hss}
\vskip -0.1truein }

\newcommand{\twofigspec}[2]
{
\hbox to\hsize{\hss
    \vbox{\psfig{figure=\mydir/#1,width=2.7in,height=3.6in}}\qquad
    \vbox{\psfig{figure=\mydir/#2,width=2.7in,height=3.6in}}
    \hss}
\vskip -0.1truein
\hbox to\hsize{\hss
    \vbox{ \begin{center}\mbox{\footnotesize \hspace{0.0in} {(a)}
                     \hspace{2.5in} {(b)}  }  \end{center} }
    \hss}
}
\newcommand{\twofigland}[2]
{
\hbox to\hsize{\hss
    \vbox{\psfig{figure=\mydir/#1,width=2.8in,height=1.8in}}\qquad
    \vbox{\psfig{figure=\mydir/#2,width=2.8in,height=1.8in}}
    \hss}
\vskip -0.1truein
\hbox to\hsize{\hss
    \vbox{ \begin{center}\mbox{\footnotesize \hspace{0.1in} {(a)}
                     \hspace{2.8in} {(b)}  }  \end{center} }
    \hss}
}
\newcommand{\twofigl}[2]
{
\hbox to\hsize{\hss
    \vbox{\psfig{figure=\mydir/#1,width=1.9in,height=1.9in}}\qquad
    \vbox{\psfig{figure=\mydir/#2,width=1.9in,height=1.9in}}
    \hss}
\vskip -0.1truein
\hbox to\hsize{\hss
    \vbox{ \begin{center}\mbox{\footnotesize \hspace{0.1in} {(a)}
                     \hspace{1.9in} {(b)}  }  \end{center} }
    \hss}
}

\newcommand{\fourfig}[4] 
{
\hbox to\hsize{\hss
    \vbox{\psfig{figure=\mydir/#1,width=3.0in,height=3.0in}}\qquad
    \vbox{\psfig{figure=\mydir/#2,width=3.0in,height=3.0in}}
    \hss}
\vskip -0.1truein
\hbox to\hsize{\hss
    \vbox{ \begin{center}\mbox{\footnotesize \hspace{0.2in} {(a)}
                     \hspace{3.0in} {(b)}  }  \end{center} }
    \hss}
\vskip 0.1truein
\hbox to\hsize{\hss
    \vbox{\psfig{figure=\mydir/#3,width=3.0in,height=3.0in}}\qquad
    \vbox{\psfig{figure=\mydir/#4,width=3.0in,height=3.0in}}
    \hss}
\vskip -0.1truein
    \vbox{ \begin{center}\mbox{\footnotesize \hspace{0.2in} {(c)}
                     \hspace{3.0in} {(d)}  }  \end{center} }
\hbox to\hsize{\hss
    \hss}
\vskip -0.1truein
}
\newcommand{\twofigv}[2]
{
\hbox to\hsize{\hss
    \vbox{\psfig{figure=\mydir/#1,width=4.2in,height=3.8in}}
    \hss}
\vskip -0.0truein
\hbox to\hsize{\hss
    \vbox{\begin{center} \footnotesize{(a)} \end{center}}
    \hss}
\vskip 0.1truein
\hbox to\hsize{\hss
    \vbox{\psfig{figure=\mydir/#2,width=4.2in,height=3.8in}}
    \hss}
\vskip -0.1truein
\hbox to\hsize{\hss
    \vbox{\begin{center} \footnotesize{(b)} \end{center}}
    \hss}
\vskip -0.1truein
}
\newcommand{\twofv}[4]
{
\hbox to\hsize{\hss
    \vbox{\psfig{figure=\mydir/#1,width=#3,height=#4}}
    \hss}
\vskip -0.1truein
\hbox to\hsize{\hss
    \vbox{\begin{center} \footnotesize{(a)} \end{center}}
    \hss}
\vskip 0.1truein
\hbox to\hsize{\hss
    \vbox{\psfig{figure=\mydir/#2,width=#3,height=#4}}
    \hss}
\vskip -0.1truein
\hbox to\hsize{\hss
    \vbox{\begin{center} \footnotesize{(b)} \end{center}}
    \hss}
\vskip -0.1truein
}
\newcommand{\sixfig}[8]
{
\hbox to\hsize{\hss
    \vbox{\psfig{figure=\mydir/#1,width=#7,height=#8}}\qquad
    \vbox{\psfig{figure=\mydir/#2,width=#7,height=#8}}
    \hss}
\vskip -0.1truein
\hbox to\hsize{\hss
    \vbox{ \begin{center}\mbox{\footnotesize {(a)}
                     \hspace{#7} {(b)}  }  \end{center} }
    \hss}
\vskip 0.1truein
\hbox to\hsize{\hss
    \vbox{\psfig{figure=\mydir/#3,width=#7,height=#8}}\qquad
    \vbox{\psfig{figure=\mydir/#4,width=#7,height=#8}}
    \hss}
\vskip -0.1truein
\hbox to\hsize{\hss
    \vbox{ \begin{center}\mbox{\footnotesize  {(c)}
                     \hspace{#7} {(d)}  }  \end{center} }
    \hss}
\vskip 0.1truein
\hbox to\hsize{\hss
    \vbox{\psfig{figure=\mydir/#5,width=#7,height=#8}}\qquad
    \vbox{\psfig{figure=\mydir/#6,width=#7,height=#8}}
    \hss}
\vskip -0.1truein
\hbox to\hsize{\hss
    \vbox{ \begin{center}\mbox{\footnotesize  {(e)}
                     \hspace{#7} {(f)}  }  \end{center} }
    \hss}
}

   {\VerbatimEnvironment
    \begin{Sbox}\begin{minipage}{#1}\begin{Verbatim}}%
   {\end{Verbatim}\end{minipage}\end{Sbox}
    \setlength{\fboxsep}{8pt}\fbox{\TheSbox}}
\newcommand{\RR}{{\Bbb{R}}}         

         \newcommand{\EE}{{\Bbb{E}}}
       
         \newcommand{\PP}{{\Bbb{P}}}

\begin{document}
\def\bib{B\kern-.05em{I}\kern-.025em{B}\kern-.08em}
\def\btex{B\kern-.05em{I}\kern-.025em{B}\kern-.08em\TeX}

\title{A NOTE ON BAYESIAN WAVELET-BASED ESTIMATION OF SCALING}

\author{Minkyoung Kang}
\author{Brani Vidakovic}
\affil{{\small \emph{Georgia Institute of Technology, Atlanta, GA}}}
\date{}

\maketitle

\begin{abstract}
A number of phenomena in various fields such as geology, atmospheric sciences, economics, to list a few, can be modeled as a fractional Brownian motion indexed by Hurst exponent $H$. This exponent is related to the degree of regularity and self-similarity present in the signal, and it often captures  important characteristics useful in various applications. Given its importance, a number of methods have been developed for the estimation of the Hurst exponent. Typically, the proposed methods do not utilize prior information about scaling of a signal.
Some signals are known to possess a theoretical value of the Hurst exponent, which motivates us to propose a Bayesian approach that incorporates this information via a suitable elicited prior distribution on $H$.
 This significantly improves the accuracy of the estimation, as we demonstrate by simulations.  Moreover, the proposed method is robust to small misspecifications of the prior location. The proposed method is applied to a turbulence time series for which Hurst exponent is theoretically known by Kolmogorov's K41 theory.
\end{abstract}


\section{Introduction}
A number of signals from natural phenomena possess fractal properties such as self-similarity and regular scaling. A popular tool to model such signals is the fractional Brownian motion (fBm), which was formalized by \cite{mandelbrot1968fractional} as follows.

\begin{definition}
 Fractional Brownian motion (fBm)  is a zero mean  Gaussian process $B_H(t), \, t \geq 0, \, 0<H<1$ for which
\ba
 \mathbb{E}\left({B_H(t)B_H(s)}\right) =\frac{\sigma^2}{2} (|t|^{2H}+|s|^{2H}-|t-s|^{2H}),
\ea
for $t, s \in R.$
\end{definition}
Here $\sigma>0$ is a scale parameter and $H \in (0,1)$ is a Hurst exponent.
The regularity of a sample path of fBm is characterized by $H$, and this
descriptor can be useful in a number of applications.

Several examples in which the Hurst exponent is well localized are as follows. For locally isotropic and fully developed turbulence, Kolmogorov introduced K41 theory. Following his theory, the Hurst exponent $H$ of turbulence processes is $1/3$. For physical particles, the asymptotic behavior of some Brownian motions that interact through collisions on a real line converges to an fBm with Hurst exponent $H = 1/4$ \cite{nourdin2009asymptotic,peligrad2007fractional,swanson2011stoch}. In a study of DNA sequences, Arneodo et al. mapped nucleotide sequences onto a ``DNA walk'' and determined that non-coding regions can be well modeled by a fractional Brownian motions with a Hurst exponent close to 0.6 \cite{arneodo1996wavelet}. For atmospheric turbulence, wave fronts become fractal surfaces behaving as an fBm with Hurst parameter $H = 5/6$ once they are degraded by turbulence \cite{schwartz1994turbulence,ribak1997atmospheric,perez2004modeling}. In addition, other refined models for turbulence yield various Hurst exponent values different from $1/3,$ but instead, a value that can be estimated by the local power law \cite{nelkin1975scaling,biskamp1994cascade,horbury2008anisotropic}. Such real-life phenomena are just a few examples in which we have prior information about the Hurst exponent prior to observing the data.

Thus, we develop a Bayesian scaling estimation method with non-decimated wavelet transform (NDWT) motivated by real-life signals that are known to possess a certain theoretical degree of self-similarity.  Bayesian approaches
 have been previously employed in this context. The Hurst exponent for Gaussian data was estimated with a Bayesian model in \cite{makarava2011bayesian,benmehdi2011bayesian,conti2004bayesian}. Holan et al. \cite{holan2009bayesian} developed a hierarchical Bayesian model to estimate the parameter of stationary long-memory processes. A Baysian model for the parameter estimation of auto-regressive fractionally integrated moving average (ARFIMA) processes \cite{hosking1981fractional} are
 discussed in \cite{graves2015efficient,ravishanker1997bayesian,pai1998bayesian}. These models are based on time domain data. However, the de-correlation property of wavelet transforms facilitates a simplified model construction, and multiple wavelet-based Bayesian techniques has been developed. Based on a Bayesian approach, Vannucci and Corradi \cite{vannucci1999modeling} estimated parameters for long memory process with a recursive algorithm and Markov chain Monte Carlo (MCMC) sampling. A Baysian wavelet model for ARFIMA processes is illustrated in \cite{ko2006bayesian}.

In this paper, we estimate Hurst exponent of a fractional Brownian motion (fBm) with wavelet coefficients from non-decimated wavelet transform (NDWT) and a Bayesian approach that incorporates information about the theoretical value of Hurst exponent via the location of a prior distribution. We combine the likelihood function and the prior distribution on ($H$, $\sigma^2$) to obtain non-normalized posterior distribution. Because we want to estimate the most likely $H$ value of an input signal given prior information and wavelet coefficients, we calculate $\hat{H}$, which maximizes the non-normalized posterior distribution. This is equivalent to estimating the mode of the posterior distribution, also referred to as maximum a posteriori (MAP) estimation. In addition, MAP estimation method results in an optimization problem that can be solved in various ways and yields an estimator optimal under a zero-one loss function. We apply the proposed method to simulated signals for the estimation of Hurst exponent $H$ based on prior distributions with approximately correct mean values. The results indicate that averaged mean squared error (MSE) of estimators significantly decreases with a prior distribution with a mean that matches the value of a true Hurst exponent. Moreover, when a slightly biased mean value of a prior distribution is provided, the averaged mean squared errors of the estimators from the proposed method are still lower than those from the regression-based method.

The rest of the paper is organized as follows. The second section introduces the proposed method that estimates the Hurst exponent with a Bayesian approach. The third section presents the simulation results and compare the estimation performance of the proposed method to the traditional regression method. The fourth section illustrates an application of the proposed method to a real-life data set, a turbulence velocity signal, that is known to possess Hurst exponent $H=1/3$. The last section is devoted to the concluding remarks and a future research direction.

\section{Method}
We applied a Bayesian model to wavelet coefficients in the domain of non-decimated wavelet transforms (NDWT). In multiresolution analysis of a $m$-dimensional fBm $B_H(\mathbf{t})$ with Hurst exponent $H$, a coefficient $d_j$ from multiresolution subspace at level $j$,  is related to a coefficient $d_0$  from a subspace at level 0, as \cite{flandrin1992wavelet}
\ba
d_j \overset{d}{=} 2^{-(H+m/2)j}d_0, \; d_0 \sim N(0, \sigma^2).
\ea
As wavelet coefficients at each multiresolution subspace follow a normal distribution with  mean zero 
and common variance, an average of the squared wavelet coefficients, under the assumption of independence,   follows a chi-square distribution.  The number of degrees of this distribution  is equal to the size of the original data. Based on such properties, we establish the following lemma:\\
\begin{lemma}
\label{lm:distsqtw}
Let $y_j$ be the average of squared wavelet coefficients, $\overline{d_j^2}$, in a wavelet subspace at level j. Then the distribution of $y_j$ is
\ba
g(y_j) =  & \left(\frac{1}{\Gamma(2^{mJ-1}) }\right) \left(\frac{2^{(2H+m)j+mJ}}{2\sigma^2}\right)^{2^{mJ-1}}&
   \big(y_j\big)^{2^{mJ-1}-1} \\  &\times  \exp \Big( -\frac{2^{mJ}}{2\sigma^2} y_j2^{(2H+m)j} \Big),&
\ea
where $m$ is the dimension of the signal, $H$ is the Hurst exponent, $J$ is an integer part of $\log_2 n$, and $n$ is the size of the input signal.
\end{lemma}
\noindent
The likelihood function of $(H, \sigma^2)$ conditional on observations of averaged energies from levels $j_1, \dots , j_2$ is
\ba
{\cal  L}(H,\sigma^2 | y_{j_1}, \dots , y_{j_2} )=\prod_{i=j_1}^{j_2} g(y_i).
\ea
We use beta distribution and non-informative prior $1/\sigma^2$ as independent priors on $H$ and $\sigma^2,$ respectively,
\ba
\pi(H,\sigma^2)=\frac{H^{\alpha -1}(1-H)^{\beta-1}}{{\cal B}(\alpha,\beta)} \times \frac{1}{\sigma^2}.
\ea
The hyperparameters in beta distribution, $\alpha$ and $\beta$ are calibrated by considering the impact of effective sample size (ESS) and the mean of the beta distribution, $\frac{\alpha}{\alpha+\beta}$, which is linked to the Hurst exponent of an input signal. The ESS for the beta($\alpha, \beta$) prior is approximated with $\alpha+\beta$ and is closely related to the performance of the Bayesian estimation. For example, when ESS is large, the posterior distribution is dominated by the prior \cite{morita2008determining}. Based on simulations, we selected the ESS to be approximately 50\% the original data size, but the ESS can be calibrated based on the level of certainty about $H$. The larger the ESS is, the more confident we are about the mean of a prior, that is,  about the ``true'' value of $H$.
\begin{theorem}
The maximum a posteriori  (MAP) estimator of $H$ is a solution to the following non-linear system:
\be
 \left\{ \begin{array}{lll}
 \frac{\partial \pi(H,\sigma^2|y_{j_1},\dots , y_{j_2} )}{\partial \sigma^2}=&-\Big(\frac{bc+2}{2}\Big)\frac{1}{\sigma^2}+\frac{b}{2\sigma^4}\sum_{j=j_1}^{j_2}y_j2^{(2H+m)j}=0 \\ \\
\frac{\partial \pi(H,\sigma^2|y_{j_1},\dots , y_{j_2} )}{\partial H}=&\frac{\alpha-1}{H}-\frac{\beta-1}{1-H} +  b\ln\ln2\sum_{j=j_1}^{j_2}j\\&-\frac{\ln2\sum_{j=j_1}^{j_2}y_jj2^{(2H+m)j}}{\sum_{j=j_1}^{j_2}y_j2^{(2H+m)j }}(bc+2) =0
\end{array} \right.
\label{solution}
\ee
\end{theorem}
Details of derivation and solution of (\ref{solution}) are deferred to Appendix. As the closed form solution that satisfies the non-linear system (\ref{solution}) is not available and given that the value of $H$ ranges only from 0 to 1, we approximately solve the equations by inserting sequentially increasing $H$ from 0 to 1 with increments of $10^{-7}.$

\section{Simulations}
\begin{figure*}
\begin{center}
\vspace{-1cm}
        \subfigure[$H=0.3$]{
            \includegraphics[width=0.55\columnwidth]{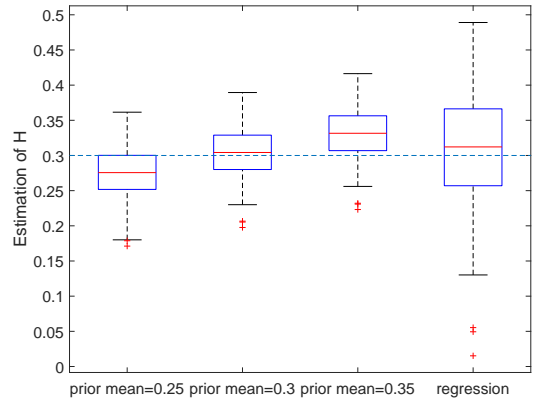}
        }
        \subfigure[$H=0.5$]{
            \includegraphics[width=0.55\columnwidth]{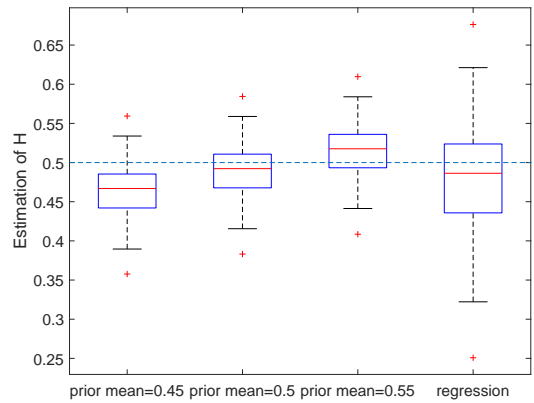}
        }
        \subfigure[$H=0.7$]{
            \includegraphics[width=0.55\columnwidth]{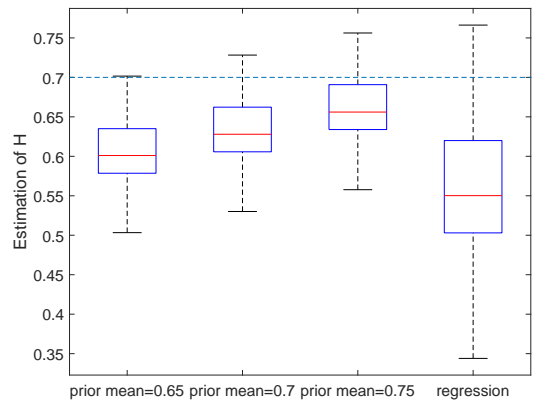}
        }
        \caption{Estimation results of simulated 200 1-D fBm's with size $2^{11}$ when Hurst exponent is 0.3, 0.5, and 0.7 under various prior settings. }
                \label{fig:simul}
\end{center}
\end{figure*}

In this section, we compare the estimation performance of the proposed method to that of non-decimated wavelet transform-based method that uses no prior information on $H$ and estimates scaling by regression, as standardly done.
 The mean, variance, mean squared error, and squared bias are reported. We simulated three sets of two hundred one-dimensional (1-D) fractional Brownian motions (fBm's) of size $2^{11}$ with Hurst exponents 0.3, 0.5, and 0.7 each. Next, we estimated the Hurst exponent of each signal using the proposed method and the traditional regression-based method. We perform an NDWT of depth 8 using Haar wavelet and analyze resulting wavelet coefficients on the $4^{th}$, $5^{th}$, and $6^{th}$ levels, noting that resolution increases with the level index and that the finest level of detail is 10. The prior distribution for $H$ is the beta with specified hyperparameters. For each set, we use three sets of prior hyperparameter settings. The prior means  are taken the same as the real (used for simulation) value, and 0.05 higher or lower than the real value, so that the effect of prior robustness can be observed. The parameters of different prior distribution settings are in Table \ref{tab:parameter}. \\
\begin{table}[h]
\begin{center}
\begin{tabular}{| c||c|c| c| c |c |c |c|c|c|c|c|} \hline
$\mu$ &  0.25 &	0.3 &	0.35 &	0.45 &	0.5 &	0.55 &	0.65 &	0.7 &	0.75 \\ \hline
$\alpha$ & 256 &	307.2 &	358.4 &	460.8 &	512 &	563.2 &	665.6 &	716.8 &	768 \\ \hline
$\beta$ &	768 &	716.8 &	665.6 &	563.2 &	512 &	460.8 &	358.4 &	307.2 &	256 \\ \hline
\end{tabular}
\caption{Setting of the parameters in the simulation study. Prior mean is $\mu$ and ($\alpha$,$\beta$) are parameters for beta prior.}
\label{tab:parameter}
\end{center}
\end{table}
Tables \ref{tb:H3}-\ref{tb:H7} summarize the estimation results in terms of mean, variance, MSE, and squared bias. Figure \ref{fig:simul} shows the estimation results as box-and-whisker plots. The proposed method yields estimators with lower MSE compared to the regression-based method under various prior settings. The estimation performance is robust to slight deviations in parameters  of the prior.   Even if the mean of a prior  differs from the value of a true Hurst exponent, estimation performance is better than the regression-based method. Correct prior mean settings significantly enhance  the estimation performance.
 We noticed, that due to autocorrelations among the NDWT wavelet coefficients,  regression-based scaling estimation  suffers from bias for Hurst exponents exceeding 1/2.  Such bias is substantially alleviated by the proposed method.

\begin{table}
\centering
\begin{subtable}
\centering
\begin{tabular}{| l | c | c |  c  |  c |} \hline
   & \multicolumn{3}{|c|}{Prior mean} & Regression \\
   \cline{2-4}
   & 0.25 & 0.3 & 0.35 &  \\ \hline \hline
Mean &  	0.2756 & 	0.3043 & 	0.3316 & 	0.3100 \\ \hline
Variance & 	 	0.0013 & 	0.0013 & 	0.0013 & 	0.0068 \\ \hline
MSE &  	0.0018 &	0.0013 & 0.0023 &	0.0068 \\ \hline
Squared bias  & 	0.0006 & 1.45E-5 & 	0.0010 & 	1.71E-5  \\ \hline
\end{tabular}
\caption{Estimation performance comparison under various prior settings with simulated 200 1-D fBms of size $2^{11}$ when Hurst exponent $H=0.3$.}
\label{tb:H3}
 \begin{tabular}{| l | c | c |  c  |  c |} \hline
   & \multicolumn{3}{|c|}{Prior mean} &  Regression  \\
   \cline{2-4}
   & 0.45 & 0.5 & 0.55 &  \\ \hline \hline
Mean &  	0.4669 & 	0.4922 & 	0.5176 & 	0.4863 \\ \hline
Variance &  0.0010 &  	0.0010 & 	0.0010 &	0.0043 \\ \hline
MSE & 	 	0.0023 & 	0.0011 & 	0.0012 & 	0.0047 \\ \hline
Squared bias &  0.0013 & 	0.0001 & 	0.0002 & 	0.0004 \\ \hline
\end{tabular}
\caption{As in Table \ref{tb:H3}, but for $H=0.5.$}
\label{tb:H5}
 \begin{tabular}{| l | c | c |  c  |  c |} \hline
   & \multicolumn{3}{|c|}{Prior mean} &   Regression   \\
   \cline{2-4}
   & 0.65 & 0.7 & 0.75 &  \\ \hline \hline
Mean & 	 	0.6280 & 	0.6561 & 	0.6858 & 	0.5502 \\ \hline
Variance &  	0.0014 & 	0.0014 & 	0.0015 & 	0.0062 \\ \hline
MSE & 	0.0059 & 0.0029 & 0.0015 & 	0.0255 \\ \hline
Squared bias & 	 	0.0045 & 	0.0015 & 	0.0001 & 	0.0193 \\ \hline
\end{tabular}
\caption{As in Table \ref{tb:H3}, but for $H=0.7.$}
\label{tb:H7}
\end{subtable}
\end{table}

\section{An Application}
\begin{figure*}
        \subfigure[$U$ velocity component of turbulence]{
            \includegraphics[width=0.45\columnwidth]{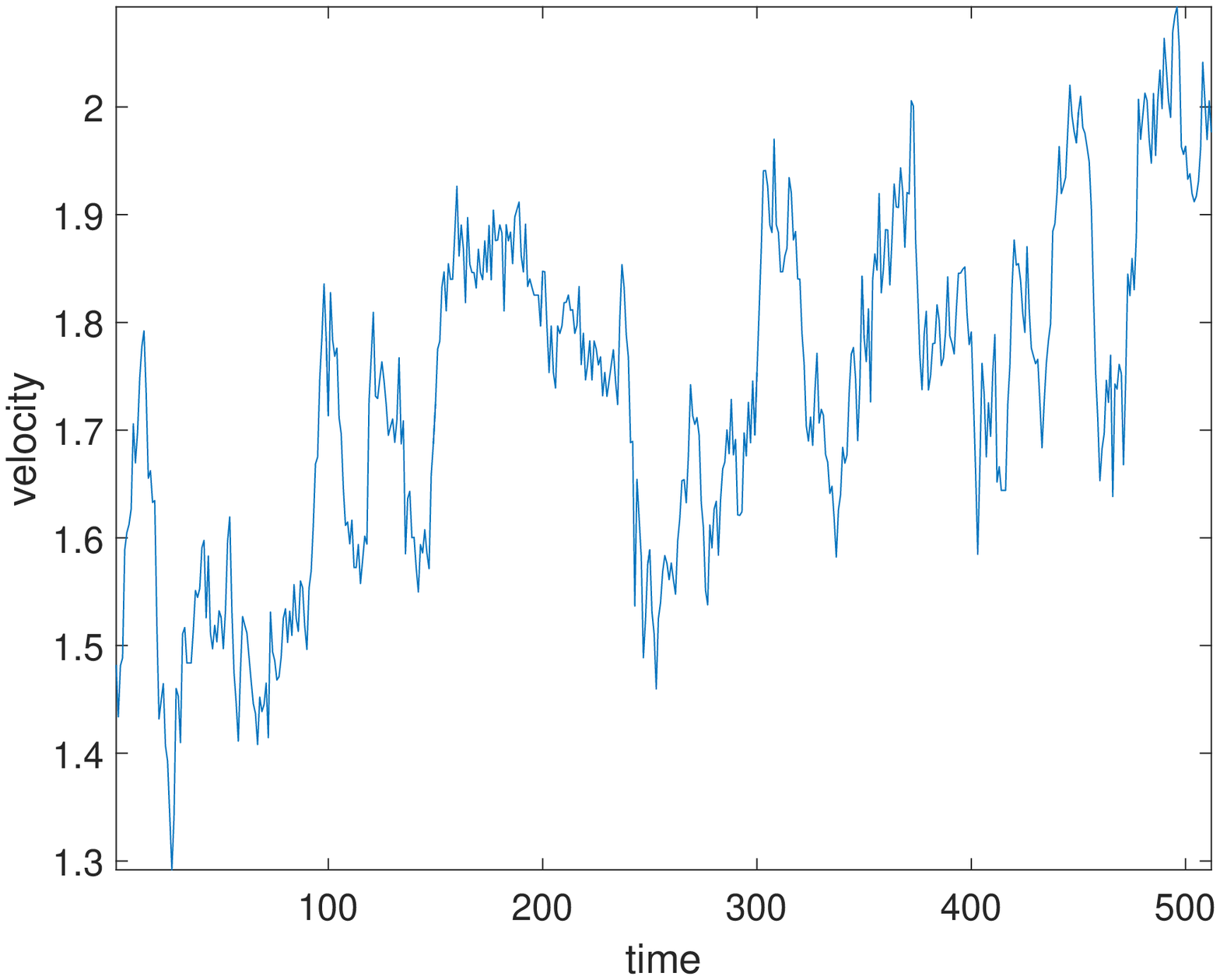}
        }
        \subfigure[Scaling behavior in the wavelet domain]{
            \includegraphics[width=0.45\columnwidth]{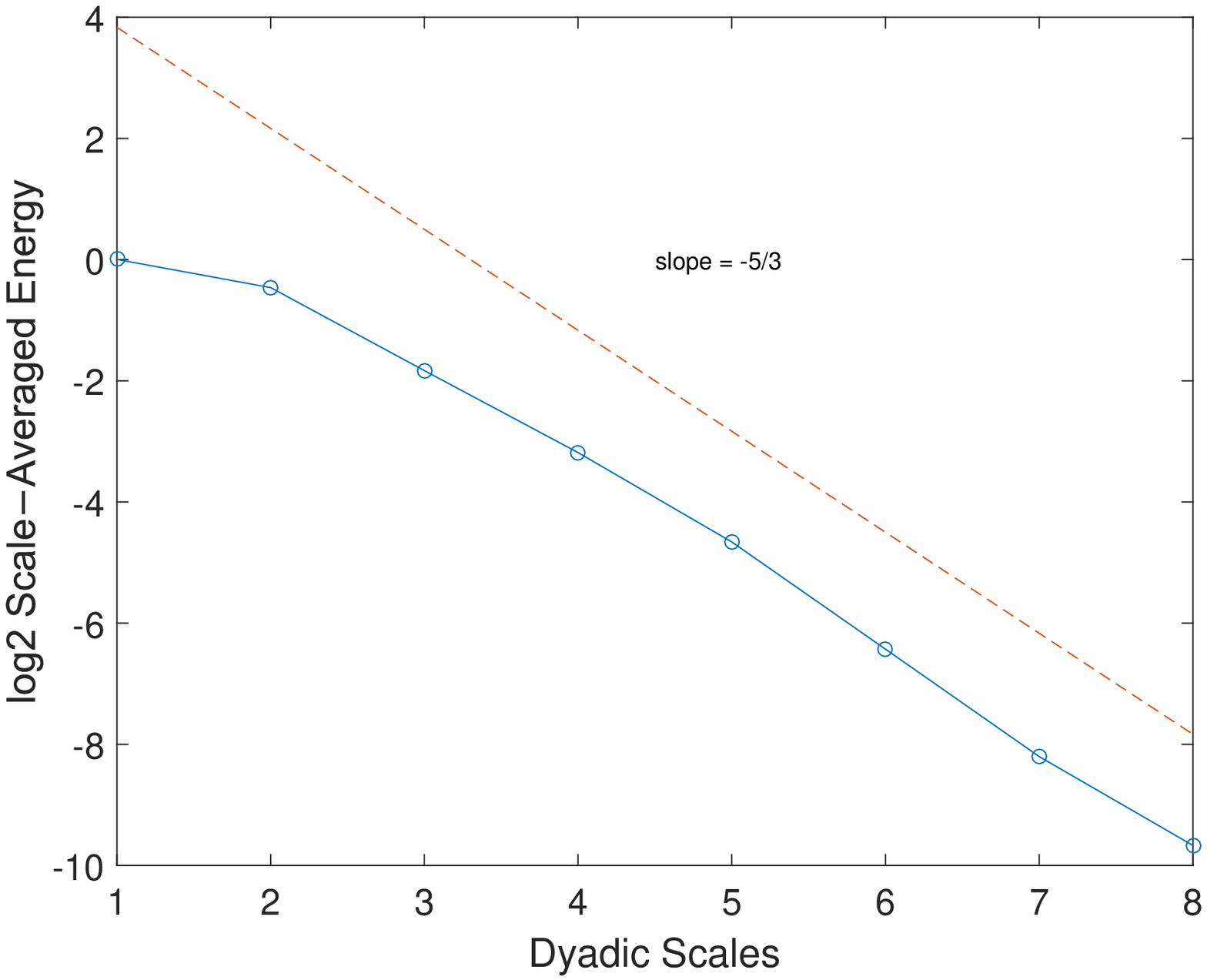}
        }
        \caption{$U$ velocity component of turbulence in time and wavelet domains.}
                \label{fig:uvel}
\end{figure*}

As an example with a real-life measurements that scale, we apply the proposed method to a dataset that traces the velocity
components of turbulence. Measurements are taken with sampling frequency $(f_s)$ of 56 Hz and period $(T_p)$ of 19.5 minutes at Duke Forrest (Durham, NC) on July 12, 1995. The data set was from a triaxial sonic anemometer (Gill Instruments/1012R2) mounted on a mast 5.2 $m$ above the ground surface over an {\it Alta Fescue} grass site. We select the $U$ component of the velocity with size $2^9$ and use it to compare the estimators from the  proposed   and regression-based methods. Based on Kolmogorov's {\bf K41} theory, we know that measurements of velocity components should have Hurst exponent close to $H=1/3$. Therefore, for the proposed method, we set the prior distribution to be the beta distribution with parameters, $\alpha=85.3$ and $\beta=170.7$, which is apriori centered at 1/3. We perform NDWT of depth 8 on the input signal and use wavelet coefficients from the eighth to the fifth level for calculations in both methods. We obtain $\hat{H}=0.341$ with the regression-based method while $\hat{H}=0.335$ with the proposed method. Figure \ref{fig:uvel} depicts the input turbulence signal in time domain and its wavelet spectrum by an NDWT.

\section{Conclusions}
A theoretical value of Hurst exponent $H$ is available for some signals, but standard scaling estimation methods do not utilize such information. We proposed a Bayesian scaling estimation method that incorporates theoretical scaling information via a prior distribution  and estimates $H$ with a MAP principle. The proposed method yields lower mean squared errors in simulations, and such performance was robust to small misspecification in the prior location. The method applied to a turbulence velocity signal yields an estimator of $H$ close to the theoretical value.

\section{Appendix}
Let $d_j = d_{j{\bm k}}$ be an arbitrary (w.r.t. $\bm{k}$) wavelet coefficient from the $j^{th}$ level of the non-decimated wavelet decomposition of the $m$-dimensional fractional Brownian motion $B_H(\omega, \bm{t}), \bm t \in \Bbb{R}^m$,
\ba
d_j = \int_{\RR^m}~ B_H(\omega, \bm t) \psi^*_{j\bm{k}}(\bm{t}) dt, \mbox{~~for some fixed $\bm{k}=
(k_1, \dots, k_m)$}
\ea
Here $\psi^*_{j\bm{k}}(\bm{t}) = \prod_{i=1}^k \psi^*_{jk_i}(t_i)$ where $\psi^*$ is either $\psi$ or $\phi$, but in the product there is at least one $\psi.$
It is known that \cite{flandrin1992wavelet}
\ba
d_j \stackrel{d}{=} 2^{-(H + m/2) j } ~d_0,
\ea
where $d_0$ is a coefficient from the level $j=0,$ and $\stackrel{d}{=}$
means equality in distributions.

Coefficient $d_j$ is a random variable with expectation
\ba
\mathbb{E}  d_j = 0, ~~~\var d_j = \mathbb{E} d_j^2 =  2^{-(2 H + m) j } ~\sigma^2,
\ea
where $\sigma^2 =   \var d_0^2.$

The fBm $B_H(\bm t)$ is a Gaussian $m$-dimensional field, thus
\ba
d_j \sim {\cal N}(0, 2^{-(2 H + m )j} \sigma^2 ).
\ea

The rescaled  ``energy'' is
\ba
 \frac{2^{(2 H + m )j} }{\sigma^2}  d_j^2 \sim \chi^2_1
 \ea
while, under assumption of independence,
\ba
 \frac{ 2^{(2 H + m )j} }{\sigma^2} \sum_{\bm k \in
 \mbox{\footnotesize $j$th level}} d_{j\bm k}^2
 =  \frac{2^{(2 H + m)j +mJ} }{\sigma^2} ~ \overline{d_{j}^2}
\ea
has $\chi^2_{2^{mJ}}$ distribution. Here $J$ is the integer part of
the logarithm for base 2 of the size of the signal.

Here, $\overline{d_{j}^2}$ is the average energy in $j^{th}$ level.

Thus,
\ba
\overline{d_{j}^2} \stackrel{d}{=}  2^{-(2 H +  m )j - m J } \sigma^2 \chi^2_{2^{m J}}.
\ea

From this,

\ba
\EE \overline{d_{j}^2} = \sigma^2 2^{-(2 H + m)j - m J } \EE \chi^2_{2^{m J}} =
 2^{-(2 Hj + mj) } \sigma^2,
\ea
and
\ba
\var \overline{d_{j}^2} = \sigma^4 2^{-(4 H + 2 m)j - 2 m J } \times 2 \cdot 2^{m J} =
 2^{-4 H j - 2 m j - m J + 1} \sigma^4.
\ea

The density of
$\overline{d_{j}^2}$ for fixed $H, j, m,$ and $\sigma^2$ is
\ba
\label{eq:dens}
g(y_j)  =& \left(\frac{1}{\Gamma(2^{mJ-1}) }\right) \left(\frac{2^{(2H+m)j+mJ}}{2\sigma^2}\right)^{2^{mJ-1}}&
    \big(y_j\big)^{2^{mJ-1}-1} \\  & \times   \exp\Big(-\frac{2^{mJ}}{2\sigma^2} y_j2^{(2H+m)j} \Big).&
\ea


Indeed, the cdf of $\overline{d_{j}^2}$ is
\ba
G(y_j)=\PP( \overline{d_{j}^2} \leq y_j ) = \PP\left(\chi^2_{2^{mJ}}
\leq \frac{2^{(2H+m)j+mJ}}{\sigma^2}y_j \right).
\ea
Then,
\ba
g(y) = G'(y) = f(h(y))~ |h'(y)|,
\ea
with $h(y_j) = \frac{2^{(2H+m)j+mJ}}{\sigma^2}y_j$ and
$f(x) = \frac{1}{ 2^{n/2} \Gamma(n/2)} x^{n/2-1} \exp \{-x/2\}, ~x \geq 0, \mbox{ ~~for $n = 2^{mJ}$}.$
Once the energy at each level $j$, $y_j$, is calculated, we can obtain the likelihood:
\begin{align*}
&{\cal L}(H, \sigma^2| y_{j_1},\dots, y_{j_2})  =\prod_{j=j_1}^{j_2} g(y_j) =  \left(\frac{(2\sigma^2)^{-2^{mJ-1} }}{\Gamma(2^{mJ-1}) }\right)^{(j_2-j_1+1)} \times \\ &\prod_{j=j_1}^{j_2} \left( {2^{(2H+m)j+mJ}} \right)^{2^{mJ-1}}
  \big(y_j\big)^{2^{mJ-1}-1} \times \exp\Big(-\sum_{j=j_1}^{j_2}\frac{2^{mJ}}{2\sigma^2} y_j2^{(2H+m)j} \Big)=  \\
   &\left(\frac{(2\sigma^2)^{-b/2 }}{\Gamma(2^{b/2}) }\right)^{c}\prod_{j=j_1}^{j_2} \left( {2^{(2H+m)j }b} \right)^{b/2}
    \big(y_j\big)^{b/2-1} \times \exp\Big(-\frac{b}{2\sigma^2}\sum_{j=j_1}^{j_2} y_j2^{(2H+m)j} \Big),
\end{align*}
where $b =2^{mJ}$ and $c=j_2-j_1+1$.

To obtain an expression proportional to the posterior distribution, we multiply likelihood function with a prior distribution, $\pi(H, \sigma^2)$,
\ba
{\cal L}(H, \sigma^2| y_{j_1},\dots, y_{j_2})
\times \pi(H, \sigma^2).
\ea
As the Hurst exponent is supported on interval $(0,1),$ we selected beta$(\alpha, \beta)$ distribution as the prior on $H$. For the prior distribution on $\sigma^2$, we selected a non-informative (improper) prior  $\frac{1}{\sigma^2}$.
The parameters $H$ and $\sigma^2$ are considered apriori independent, so their joint prior is
\ba
\pi(H,\sigma^2)=\frac{1}{\sigma^2}\frac{\Gamma(\alpha+\beta)}{\Gamma(\alpha)\Gamma(\beta)}H^{\alpha-1}(1-H)^{\beta-1}.
\ea
A non-normalized posterior is
\begin{align}
\nonumber F &= \pi(H,\sigma^2){\cal L}(H, \sigma^2|  y_{j_1},\dots, y_{j_2}) = \frac{\Gamma(\alpha+\beta)}{\Gamma(\alpha)\Gamma(\beta)} H^{\alpha-1}(1-H)^{\beta-1}\Big(\frac{1}{ \Gamma(b/2)}\Big)^{c}\times \\ &\Big(\frac{b}{2}\Big)^{bc/2} \Big(\frac{1}{ \sigma^2}\Big)^{bc/2+1} \prod_{j=j_1}^{j_2}2^{(2H+m)jb/2}\big(y_j\big)^{b/2-1}\exp\Big(-\frac{b}{2\sigma^2}\sum_{j=j_1}^{j_2}y_j2^{(2H+m)j} \Big).
\label{eq:F}
\end{align}
Taking logarithm of (\ref{eq:F}) yields
\begin{align}
\ln F  =&-\frac{bc+2}{2}\ln \sigma^2 +\sum_{j=j_1}^{j_2}\bigg[\frac{(2H+m)jb}{2}\ln2 +\frac{b-2}{2} \ln y_j \bigg] \nonumber
\\  &-\frac{b}{2\sigma^2}\sum_{j=j_1}^{j_2}y_j2^{(2H+m)j}+ \ln \bigg[ \frac{\Gamma(\alpha+\beta)}{\Gamma(\alpha)\Gamma(\beta)} \bigg] +(\alpha-1)\ln H \nonumber \\
&+(\beta-1)\ln (1-H) -c\ln \big[ \Gamma(b/2) \big] + \frac{bc}{2}\ln[b/2].
\label{eq:lnf}
\end{align}
The estimator that maximizes the posterior, maximizes its non-normalized version
as well.
First, we obtain $\sigma^2$ that maximizes the likelihood by taking derivative
\baa
\frac{\partial \ln F}{\partial \sigma^2} & = -\Big(\frac{bc+2}{2}\Big)\frac{1}{\sigma^2}+\frac{b}{2\sigma^4}\sum_{j=j_1}^{j_2}y_j2^{(2H+m)j}=0\\
\sigma^2& = \frac{b\sum_{j=j_1}^{j_2}y_j2^{(2H+m)j }}{ bc+2}
\label{eq:sig}
\end{align}
Using (\ref{eq:sig}) obtained, we express (\ref{eq:lnf}) as a function of $H,$   and take derivative to obtain $H$ that maximizes the likelihood,
\be
\ln F  =& -&\frac{bc+2}{2}\ln \bigg[\frac{b\sum_{j=j_1}^{j_2}y_j2^{(2H+m)j }}{ bc+2} \bigg] \nonumber \\ &+&\sum_{j=j_1}^{j_2}\bigg[\frac{(2H+m)jb}{2}\ln2 +\frac{b-2}{2} \ln y_j \bigg] \nonumber \\
 &-&\frac{bc+2}{2 }
 + \ln \bigg[ \frac{\Gamma(\alpha+\beta)}{\Gamma(\alpha)\Gamma(\beta)} \bigg] +(\alpha-1)\ln H +(\beta-1)\ln (1-H)\nonumber  \\ &-& c\ln \big[ \Gamma(b/2) \big] + \frac{bc}{2}\ln[b/2].\\
\frac{\partial \ln F}{\partial H}=&-&(bc+2)\frac{\ln2  \sum_{j=j_1}^{j_2}y_jj2^{(2H+m)j}}{\sum_{j=j_1}^{j_2}y_j2^{(2H+m)j}}\, +\, b\ln2\sum_{j=j_1}^{j_2}j  +\frac{\alpha-1}{H} \nonumber \\ &-& \frac{\beta-1}{1-H} =0.
\label{eq:HF}
\ee
There is no closed form solution for $H$, so we numerically approximate its value by solving equations in (\ref{eq:HF}).

\bibliography{TB}

\begin{thebibliography}{10}

\bibitem{arneodo1996wavelet}
A~Arneodo, Y~d'Aubenton Carafa, E~Bacry, PV~Graves, JF~Muzy, and C~Thermes.
\newblock Wavelet based fractal analysis of {DNA} sequences.
\newblock {\em Physica D: Nonlinear Phenomena}, 96(1):291--320, 1996.

\bibitem{benmehdi2011bayesian}
S~Benmehdi, N~Makarava, N~Benhamidouche, and M~Holschneider.
\newblock {B}ayesian estimation of the self-similarity exponent of the {N}ile
  river fluctuation.
\newblock {\em Nonlinear Processes in Geophysics}, 18(3):441--446, 2011.

\bibitem{biskamp1994cascade}
D~Biskamp.
\newblock Cascade models for magnetohydrodynamic turbulence.
\newblock {\em Physical Review E}, 50(4):2702, 1994.

\bibitem{conti2004bayesian}
PL~Conti, A~Lijoi, and F~Ruggeri.
\newblock A {B}ayesian approach to the analysis of telecommunication systems
  performance.
\newblock {\em Applied Stochastic Models in Business and Industry},
  20:305--321, 2004.

\bibitem{flandrin1992wavelet}
Patrick Flandrin.
\newblock Wavelet analysis and synthesis of fractional {B}rownian motion.
\newblock {\em Information Theory, IEEE Transactions on}, 38(2):910--917, 1992.

\bibitem{graves2015efficient}
T~Graves, RB~Gramacy, CLE Franzke, and NW~Watkins.
\newblock Efficient {B}ayesian inference for {ARFIMA} processes.
\newblock {\em Nonlinear Processes in Geophysics Discussions}, 2:573--618,
  2015.

\bibitem{holan2009bayesian}
Scott Holan, Tucker McElroy, Sounak Chakraborty, et~al.
\newblock A {B}ayesian approach to estimating the long memory parameter.
\newblock {\em {B}ayesian Analysis}, 4(1):159--190, 2009.

\bibitem{horbury2008anisotropic}
Timothy~S Horbury, Miriam Forman, and Sean Oughton.
\newblock Anisotropic scaling of magnetohydrodynamic turbulence.
\newblock {\em Physical Review Letters}, 101(17):175005, 2008.

\bibitem{hosking1981fractional}
Jonathan~RM Hosking.
\newblock Fractional differencing.
\newblock {\em Biometrika}, 68(1):165--176, 1981.

\bibitem{ko2006bayesian}
Kyungduk Ko and Marina Vannucci.
\newblock {B}ayesian wavelet analysis of autoregressive fractionally integrated
  moving-average processes.
\newblock {\em Journal of Statistical Planning and Inference},
  136(10):3415--3434, 2006.

\bibitem{makarava2011bayesian}
Natallia Makarava, Sabah Benmehdi, and Matthias Holschneider.
\newblock {B}ayesian estimation of self-similarity exponent.
\newblock {\em Physical Review E}, 84(2):021109, 2011.

\bibitem{mandelbrot1968fractional}
Benoit~B Mandelbrot and John~W Van~Ness.
\newblock Fractional {B}rownian motions, fractional noises and applications.
\newblock {\em SIAM review}, 10(4):422--437, 1968.

\bibitem{morita2008determining}
Satoshi Morita, Peter~F Thall, and Peter M{\"u}ller.
\newblock Determining the effective sample size of a parametric prior.
\newblock {\em Biometrics}, 64(2):595--602, 2008.

\bibitem{nelkin1975scaling}
Mark Nelkin.
\newblock Scaling theory of hydrodynamic turbulence.
\newblock {\em Physical Review A}, 11(5):1737, 1975.

\bibitem{nourdin2009asymptotic}
Ivan Nourdin, Anthony R{\'e}veillac, et~al.
\newblock Asymptotic behavior of weighted quadratic variations of fractional
  {B}rownian motion: the critical case {$H$}= 1/4.
\newblock {\em The Annals of Probability}, 37(6):2200--2230, 2009.

\bibitem{pai1998bayesian}
Jeffrey~S Pai and Nalini Ravishanker.
\newblock {B}ayesian analysis of autoregressive fractionally integrated
  moving-average processes.
\newblock {\em Journal of Time Series Analysis}, 19(1):99--112, 1998.

\bibitem{peligrad2007fractional}
Magda Peligrad and Sunder Sethuraman.
\newblock On fractional {B}rownian motion limits in one dimensional
  nearest-neighbor symmetric simple exclusion.
\newblock {\em arXiv preprint arXiv:0711.0017}, 2007.

\bibitem{perez2004modeling}
Dar{\i}o~G P{\'e}rez, Luciano Zunino, and Mario Garavaglia.
\newblock Modeling turbulent wave-front phase as a fractional {B}rownian
  motion: a new approach.
\newblock {\em JOSA A}, 21(10):1962--1969, 2004.

\bibitem{ravishanker1997bayesian}
Nalini Ravishanker and Bonnie~K Ray.
\newblock {B}ayesian analysis of vector {ARFIMA} processes.
\newblock {\em Australian Journal of Statistics}, 39(3):295--311, 1997.

\bibitem{ribak1997atmospheric}
Erez Ribak.
\newblock Atmospheric turbulence, speckle, and adaptive optics.
\newblock {\em Annals of the New York Academy of Sciences}, 808(1):193--204,
  1997.

\bibitem{schwartz1994turbulence}
C~Schwartz, G~Baum, and EN~Ribak.
\newblock Turbulence-degraded wave fronts as fractal surfaces.
\newblock {\em JOSA A}, 11(1):444--451, 1994.

\bibitem{swanson2011stoch}
Jason Swanson.
\newblock Fluctuations of the empirical quantiles of independent {B}rownian
  motions.
\newblock {\em Stochastic Processes and their Applications}, 2011.

\bibitem{vannucci1999modeling}
Marina Vannucci and Fabio Corradi.
\newblock Modeling dependence in the wavelet domain.
\newblock In {\em {B}ayesian inference in wavelet-based models}, pages
  173--186. Springer, 1999.

\end{thebibliography}

\end{document}